\documentclass[twocolumn,prd, superscriptaddress, showpacs, floatfix]{revtex4-1}
\usepackage{hyperref,xcolor}
\usepackage{amsmath,amssymb}
\usepackage{graphicx}
\usepackage{xcolor}
\usepackage{bm}
\usepackage{layouts}
\usepackage{pgf,tikz}
\usetikzlibrary{arrows}
\usetikzlibrary[patterns]
\definecolor{zzttqq}{rgb}{0.6,0.2,0}
\definecolor{cqcqcq}{rgb}{0.75,0.75,0.75}

\newcommand{\R}{\ensuremath{\mathbb{R}}}

\newcommand{\rd}{\ensuremath{\mathrm{d}}}

\providecommand{\abs}[1]{\lvert#1\rvert}

\newcommand{\expv}[1]{\left\langle#1\right\rangle}

\newcommand{\be}{\begin{equation}}
\newcommand{\ee}{\end{equation}}
\newcommand{\benn}{\nonumber\begin{equation}}
\newcommand{\eenn}{\nonumber\end{equation}}
\def\bea{\begin{eqnarray}} \def\eea{\end{eqnarray}}
\def\beann{\begin{eqnarray*}} \def\eeann{\end{eqnarray*}}
\def\lsim{\raise0.3ex\hbox{$<$\kern-0.75em\raise-1.1ex\hbox{$\sim$}}}
\def\gsim{\raise0.3ex\hbox{$>$\kern-0.75em\raise-1.1ex\hbox{$\sim$}}}

\graphicspath {{figures/}}

\begin{document}

\title{{\hfill \texttt{\footnotesize CERN-PH-TH-2015-110}}
\vfill
U(1) lattice gauge theory with a topological action}
\date{\today}
\author{Oscar Akerlund}
\affiliation{Institut f\"ur Theoretische Physik, ETH Z\"urich, CH-8093 Z\"urich, Switzerland}
\author{Philippe de Forcrand}
\affiliation{Institut f\"ur Theoretische Physik, ETH Z\"urich, CH-8093 Z\"urich, Switzerland}
\affiliation{CERN, Physics Department, TH Unit, CH-1211 Geneva 23, Switzerland}

\begin{abstract}
We investigate the phase diagram of the compact $U(1)$ lattice gauge theory in four dimensions using a non-standard
action which is invariant under continuous deformations of the plaquette angles.
Just as for the Wilson action, we find a weakly first order transition,
separating a confining phase where magnetic monopoles condense, and a 
Coulomb phase where monopoles are dilute. 
We also find a third phase where monopoles are completely absent.
The topological action offers an algorithmic advantage for the computation
of the free energy.
\end{abstract}

\maketitle

\section{Introduction}\label{sec:introduction}\noindent
In recent years lattice Quantum Field Theory has seen a surge of
efforts to construct new lattice actions which aim at improving
the approach to the continuum limit. The best-known strategy
is that advocated by Symanzik, where irrelevant operators of 
higher and higher dimension are added to the ``standard''
(e.g.~Wilson plaquette) action, with coefficients adjusted perturbatively
or non-perturbatively to cancel discretization errors of
the corresponding power in the lattice spacing $a$~\cite{Symanzik:1983gh,Weisz:1982zw}.
This kind of improvement is thus a parametric one, allowing
for a faster approach to the continuum limit than exhibited by
the ``standard'' action.

However, this is not the only possible strategy for improvement.
It has long been recognized that departure from the continuum
limit is more violent for large fields, so that suppressing
these large fields produces a non-parametric improvement~\cite{Meurice:2001ir}.
For instance, this happens when one trades the Wilson
action for the Manton action~\cite{Michael:1988nd}, 
based on the length of the 
geodesic in group space, or for a ``perfect'' action~\cite{Rufenacht:2001pi}: 
large fields, corresponding to 
small values of the plaquette trace, are more suppressed
than with the Wilson action, and at the same time continuum
behavior is better approximated for a given value of
the lattice spacing $a$.

A more radical suppression of large fields is achieved by imposing
a strict cutoff: for instance, in a spin model one can demand
that neighboring spin angles do not differ by more than a limiting
value; or in a gauge theory, one may require that the plaquette
trace be larger than a limiting value. The best-known example
of the latter is the positive-plaquette action for $SU(2)$ 
lattice gauge theory~\cite{Mack:1981gy,Bornyakov:1991gq,Fingberg:1994ut}. While the approach to
the continuum limit is also improved in this strategy,
an important side-effect may happen. Localized topological
defects can only form if the cutoff is not too restrictive.
For instance, an $O(2)$ spin model on a square lattice can
support vortices only if the spins can rotate by $\pi/2$ or more
between neighboring sites. If not, the disordered phase of
this system disappears entirely. Thus, the cutoff may change
the phase diagram of the model. A similar situation occurs
in lattice gauge theory: as pointed out by L\"uscher~\cite{Luscher:1981zq},
if the plaquette trace is restricted to ``admissible'' values greater 
than about 0.97 (for $SU(2)$), changes in the topological charge become 
impossible, and topology becomes well defined on the lattice. 
Topological sectors arise as in the continuum theory.

Here, we consider the extreme strategy where the action consists
{\em only} of a cutoff. In other words, the action takes only
two values: 0 if all cutoff restrictions are satisfied, $+\infty$
if not. This kind of action has been called {\em topological}~\cite{Bietenholz:2010xg},
because it does not have any classical small-$a$ limit, and the
action remains invariant under small admissible deformations of the field.
A simple example of topological action for an $O(N)$ spin model is:

\begin{equation}
  \label{eq:act_spin}
  S = \sum_{\expv{i,j}}R_\theta(S_i\cdot{}S_j),\;\; R_\theta(x) = \begin{cases}0 &x > \cos\theta\\+\infty & \text{else}
  \end{cases}.
\end{equation}

Topological actions raise an interesting puzzle: as the constraint 
between neighboring spins becomes more restrictive, the correlation length
increases and diverges; but what is the action associated with this continuum limit?
Several studies have investigated different spin models~\cite{Bietenholz:2010xg,Bietenholz:2012ud}, and it has
been shown in analytically solvable $O(N)$ models that the continuum limit
is that associated with the usual, sigma-model action. In higher dimensions
numerical investigations also support this claim very strongly.

Here we want to investigate the properties of a topological action
in a gauge theory, and consider the simplest case, namely compact $U(1)$ lattice
gauge theory in 4 dimensions. Aside from the continuum limit, we also
want to study the phase diagram of this system. With the Wilson action,
a first-order phase transition separates a strong-coupling, confining phase 
and a weak-coupling Coulomb phase. This phase transition is associated
with condensation of magnetic monopoles in the strong-coupling phase~\cite{DeGrand:1980eq}.
With a topological action, the constraint on the plaquette trace, when
restrictive enough, is going to make it impossible for magnetic monopoles
to exist. This may completely alter the phase diagram of the theory.

Finally, topological actions may be interesting for algorithmic reasons:
it may be computationally easier to move in the space of admissible configurations 
since they all have the same action. While this promise has not yet been realized
for the Monte Carlo update of such configurations, in spin models or in
the gauge theory we study, we show below that extracting the free energy
(or equivalently here, the entropy) is extremely simple numerically, and
yields valuable information.

Our paper is organized as follows: we discuss the topological action of
our model in Sec.~II, the consequences for magnetic monopoles in Sec.~III,
the helicity modulus in Sec.~IV, propose some arguments about the 
continuum limit in Sec.~V, and discuss how to obtain the free energy 
in Sec.~VI. Our results on the phase diagram are presented in Sec.~VII,
followed by conclusions.

\section{The action}
\begin{figure}[htp]
\begin{tikzpicture}[line cap=round,line join=round,>=triangle 45,x=0.7cm,y=0.7cm]
\clip(-7.01,-3.01) rectangle (5,4.01);
\draw(-4,0) circle (1.4cm);
\draw [shift={(-4,0)},pattern color=zzttqq,fill=zzttqq,pattern=north east lines]  (0,0) --  plot[domain=2.76:3.51,variable=\t]({1*2*cos(\t r)+0*2*sin(\t r)},{0*2*cos(\t r)+1*2*sin(\t r)}) -- cycle ;
\draw [shift={(-4,0)},pattern color=zzttqq,fill=zzttqq,pattern=north east lines]  (0,0) --  plot[domain=4.53:5.28,variable=\t]({1*2*cos(\t r)+0*2*sin(\t r)},{0*2*cos(\t r)+1*2*sin(\t r)}) -- cycle ;
\draw [shift={(-4,0)},pattern color=zzttqq,fill=zzttqq,pattern=north east lines]  (0,0) --  plot[domain=0.13:0.88,variable=\t]({1*2*cos(\t r)+0*2*sin(\t r)},{0*2*cos(\t r)+1*2*sin(\t r)}) -- cycle ;
\draw [->] (-4,0) -- (-2,0);
\draw [->] (-4,0) -- (-4.29,1.98);
\draw [->] (-4,0) -- (-5.73,-1.01);
\draw [shift={(-4,0)},line width=2pt]  plot[domain=0.88:2.77,variable=\t]({1*2.6*cos(\t r)+0*2.6*sin(\t r)},{0*2.6*cos(\t r)+1*2.6*sin(\t r)});
\draw [shift={(-4,0)},line width=2pt]  plot[domain=-1:0.13,variable=\t]({1*2.65*cos(\t r)+0*2.65*sin(\t r)},{0*2.65*cos(\t r)+1*2.65*sin(\t r)});
\draw [shift={(-4,0)},line width=2pt]  plot[domain=3.51:4.53,variable=\t]({1*2.56*cos(\t r)+0*2.56*sin(\t r)},{0*2.56*cos(\t r)+1*2.56*sin(\t r)});
\draw(2,0) circle (1.4cm);
\draw [->] (2,0) -- (4,0);
\draw [->] (2,0) -- (3.92,0.55);
\draw [->] (2,0) -- (3.97,-0.33);
\draw [shift={(2,0)},pattern color=zzttqq,fill=zzttqq,pattern=north east lines]  (0,0) --  plot[domain=1.03:5.25,variable=\t]({1*2*cos(\t r)+0*2*sin(\t r)},{0*2*cos(\t r)+1*2*sin(\t r)}) -- cycle ;
\draw [shift={(2,0)},pattern color=zzttqq,fill=zzttqq,pattern=north east lines]  (0,0) --  plot[domain=1.38:5.6,variable=\t]({1*2*cos(\t r)+0*2*sin(\t r)},{0*2*cos(\t r)+1*2*sin(\t r)}) -- cycle ;
\draw [shift={(2,0)},pattern color=zzttqq,fill=zzttqq,pattern=north east lines]  (0,0) --  plot[domain=0.86:5.09,variable=\t]({1*2*cos(\t r)+0*2*sin(\t r)},{0*2*cos(\t r)+1*2*sin(\t r)}) -- cycle ;
\draw [shift={(2,0)},line width=2pt]  plot[domain=-0.69:0.86,variable=\t]({1*2.66*cos(\t r)+0*2.66*sin(\t r)},{0*2.66*cos(\t r)+1*2.66*sin(\t r)});
\draw (-2.00,0.25) node[anchor=north west] {$ s_1 $};
\draw (-4.5,2.55) node[anchor=north west] {$ s_2 $};
\draw (-6.2,-0.95) node[anchor=north west] {$ s_3$};
\draw (4.,0.29) node[anchor=north west] {$ s_2 $};
\draw (3.9,-0.15) node[anchor=north west] {$ s_1 $};
\draw (3.9,0.96) node[anchor=north west] {$ s_3 $};
\draw (-4.9,3.71) node[anchor=north west] {$ \delta_\text{max}>\pi/2 $};
\draw (1.2,3.71) node[anchor=north west] {$ \delta_\text{max}<\pi/2 $};
\end{tikzpicture}
\caption{Forbidden regions (hatched areas) and allowed regions (black lines) for the angle of a link surrounded by three (the others are omitted for clarity) staples $s_i$.
  When the restriction angle $\delta_\text{max}>\pi/2$ (\emph{left panel}) the region can be disconnected whereas if it is smaller
  than $\pi/2$ (\emph{right panel}) it will always be connected. $\delta_\text{max}$ is the angle between an arrow and the edge
  of the hatched area opposite to it.}
\label{fig:allowed_angles}
\end{figure}
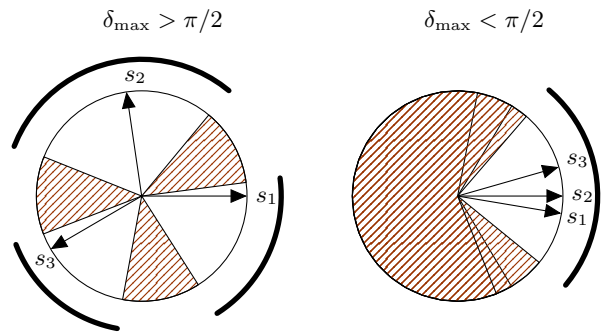

The obvious analogue of restricting the angles between neighboring spins in a spin model is to restrict the
real part of the trace of each plaquette in a gauge theory. The action then depends on one coupling $\alpha$
and is given by
\begin{equation}
  \label{eq:top_act}
  e^{-S} = \begin{cases}1 &\mathrm{ReTr}U_P > \alpha\quad \forall P\\0 & \text{else}
  \end{cases},
\end{equation}
where $P$ denotes a plaquette. Note that this formulation is independent of the gauge group but that we from
now on consider only $U(1)$ where $\mathrm{ReTr}U_P=\cos\theta_P$. We could thus equally well consider a restriction
of the plaquette angle $\theta_P$ with $\abs{\theta_P\mod2\pi}<\delta_\text{max} \equiv \arccos\alpha$.
It is also important to note that the link angles, being gauge variant, are
completely unrestricted. The most efficient way to generate configurations is to apply heatbath updates to the
links one at a time under the constraint that no plaquette angle exceeds the allowed value. In principle this is realized by just
uniformly sampling the interval $[0,2\pi]$ until an acceptable angle has been found but in some cases it might be
more efficient to explicitly construct the allowed range of values for the link to be updated. Note that a Metropolis update
based on the old value may not be ergodic since the admissible region of link angles may not be connected. See
Fig.~\ref{fig:allowed_angles}. However, there are some additional caveats to this kind of single link update which will become clear
in the discussion of the magnetic monopoles.

\section{Magnetic monopoles}
An elementary cube on the lattice contains $q$ magnetic monopoles if the outward oriented, physical
($\theta_P\in[-\pi,\pi[$) plaquette angles of its faces sum up to $2\pi q$~\cite{DeGrand:1980eq}.
It is easy to check that $q\in\{0,\pm 1,\pm 2\}$ and that a cube with $q$ monopoles must have at least one face with
physical plaquette angle $\abs{\theta_P}\geq \abs{q}\pi/3$. This immediately tells us that for $\delta_\text{max}<\pi/3$
there cannot be any monopoles and the topological action does not describe the same (lattice) physics as the Wilson
action~\footnote{However, it should also be noted that the $U(1)$ monopole is a lattice artifact which disappears in
the continuum limit also for the Wilson action.}. In fact, a change of variables from link to rescaled plaquette angles
$\theta_P/\delta_{\text{max}}$ can be used to see that all $\delta_\text{max}<\pi/3$ are equivalent up to trivial rescalings.
Let us therefore concentrate on angles larger than that.

One might think that if there is a deconfinement transition at some restriction angle $\delta_\text{max}$ then it should be at
$\delta_\text{max}=\pi/3$ since
this angle separates the region of no monopoles from a region with monopoles. This turns out to be wrong. In a sense
this is analogous to the situation with the Wilson action. At the deconfinement transition the monopole density jumps down,
but it does not jump to zero. The system can sustain a small density of monopoles without being confining. The same happens for
the topological action with a deconfinement transition at a significantly larger restriction angle than $\pi/3$.
Still, there is a non-analyticity in the monopole density at $\delta_\text{max}=\pi/3$,
which we investigate further in Sec.~VII (see Figs.~\ref{fig:dens_helicity} and \ref{fig:nmonop_top}).

\subsection{Creating monopoles}
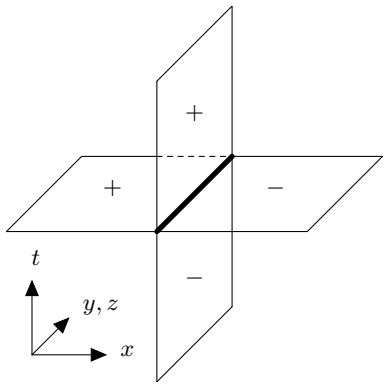
\begin{figure}[htp]
\begin{tikzpicture}[line cap=round,line join=round,>=triangle 45,x=1.0cm,y=1.0cm]
\clip(1,-1) rectangle (6,4);
\draw (3,3)-- (3,1);
\draw (3,3)-- (4,4);
\draw (4,4)-- (4,2);
\draw [line width=2pt] (3,1)-- (4,2);
\draw (3,1)-- (3,-1);
\draw (3,-1)-- (4,0);
\draw (4,0)-- (4,2);
\draw (3,1)-- (5,1);
\draw (4,2)-- (6,2);
\draw (5,1)-- (6,2);
\draw (3,1)-- (1,1);
\draw (1,1)-- (2,2);
\draw (2,2)-- (3,2);
\draw [dash pattern=on 2pt off 2pt] (3,2)-- (4,2);
\draw (2.17,1.8) node[anchor=north west] {$\mathbf{+}$};
\draw (3.26,2.8) node[anchor=north west] {$\mathbf{+}$};
\draw (4.34,1.8) node[anchor=north west] {$\mathbf{-}$};
\draw (3.26,0.6) node[anchor=north west] {$\mathbf{-}$};
\draw (1.22,0.86) node[anchor=north west] {$t$};
\draw [->] (1.34,-0.64) -- (2.34,-0.64);
\draw [->] (1.34,-0.64) -- (1.34,0.36);
\draw [->] (1.34,-0.64) -- (1.84,-0.14);
\draw (2.4,-0.4) node[anchor=north west] {$x$};
\draw (1.9,0.21) node[anchor=north west] {$y, z$};
\end{tikzpicture}
\caption{The smallest possible nontrivial loop of monopoles world lines which has Euclidean length $2\sqrt{2}a$. The $y$ and
$z$ dimensions are collapsed into one so that each cube is represented by a plaquette and each plaquette by a link.
The fat link represents the plaquette shared by all four cubes which contain a monopole. A $+(-)$ in a plaquette
symbolizes a positively(negatively) charged monopole in the corresponding cube.}
\label{fig:monop_loop}
\end{figure}

To study how the monopoles depend on the cutoff angle $\delta_\text{max}$ it is important to understand what the lowest
monopole excitation is. It is well known that every monopole is connected to an anti-monopole via a Dirac string
and that the monopole worldlines must form closed loops on the dual lattice. The shortest such loop has four vertices and Euclidean length
$2\sqrt{2}a$ where $a$ is the lattice spacing, and the smallest excitation is thus two monopoles and two anti-monopoles
each located in one of the four cubes sharing a single plaquette. See Fig.~\ref{fig:monop_loop} for an illustration.

\begin{figure}[htbp]
\begin{tikzpicture}[line cap=round,line join=round,>=triangle 45,x=0.5cm,y=0.5cm]
\clip(0,-10.5) rectangle (16,4.5);
\draw (2,2)-- (2,-2);
\draw (2,-2)-- (6,-2);
\draw (2,2)-- (6,2);
\draw (6,-2)-- (6,2);
\draw (2,2)-- (4,4);
\draw (4,4)-- (8,4);
\draw (6,2)-- (8,4);
\draw (8,4)-- (8,0);
\draw (8,0)-- (6,0);
\draw [dash pattern=on 3pt off 3pt] (6,0)-- (4,0);
\draw [dash pattern=on 3pt off 3pt] (4,0)-- (2,-2);
\draw [dash pattern=on 3pt off 3pt] (4,0)-- (4,4);
\draw (8,-2)-- (12,-2);
\draw (8,-2)-- (10,0);
\draw (10,0)-- (12,0);
\draw [dash pattern=on 3pt off 3pt] (12,0)-- (14,0);
\draw [line width=2.8pt] (12,-2)-- (14,0);
\draw (12,-2)-- (12,2);
\draw (12,2)-- (14,4);
\draw (14,4)-- (14,0);
\draw (4.54,1.34) node[anchor=north west] {$\pi$};
\draw (10.57,1.64) node[anchor=north west] {$\pm\pi$};
\draw (1,-10)-- (1,-6);
\draw (5,-6)-- (1,-6);
\draw (1,-6)-- (3,-4);
\draw (3,-4)-- (7,-4);
\draw (7,-4)-- (5,-6);
\draw (1,-10)-- (3,-8);
\draw (3,-8)-- (7,-8);
\draw (7,-4)-- (7,-8);
\draw (3,-8)-- (3,-6);
\draw [dash pattern=on 3pt off 3pt] (3,-4)-- (3,-6);
\draw (9,-10)-- (13,-10);
\draw [line width=2.8pt] (13,-6)-- (13,-10);
\draw (9,-6)-- (9,-10);
\draw (9,-6)-- (13,-6);
\draw (13,-6)-- (15,-4);
\draw (15,-4)-- (15,-8);
\draw [line width=2.8pt] (15,-8)-- (13,-10);
\draw [dash pattern=on 3pt off 3pt] (9,-10)-- (11,-8);
\draw [dash pattern=on 3pt off 3pt] (11,-8)-- (15,-8);
\draw (4.23,-6.39) node[anchor=north west] {$\pi$};
\draw (11.27,-4.56) node[anchor=north west] {$\pm\pi$};
\end{tikzpicture}
\caption{Monopole creation with a single link update (\emph{upper panel}) and a multiple link update (\emph{lower
panel}). The fat links are the ones updated and the flux of $\pi$ is spread over the plaquettes on the right which
means that the single link update is ergodic down to $\delta_{max}=\pi/2$ and the two-link one to $\pi/3$.}
\label{fig:monop_update}
\end{figure}
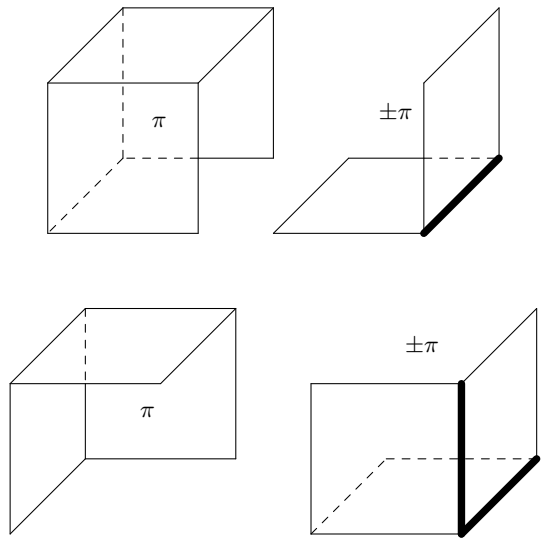

It is also important to consider how such a configuration is created from a configuration with zero monopoles.
In order to create a monopole in a given cube we need to change its flux by $2\pi$ at the same time as we
respect the constraints on the plaquette angles. It is therefore relevant to investigate the smallest constraint
angle for which a change of $2\pi$ in the flux is possible. If we update a single edge of a cube we will change
two of its six plaquettes. The sum of these changes must be $2\pi$ and the required angles can be minimized by
letting the change be distributed equally over all involved plaquettes. Hence, the restriction on the plaquette
angles gives $\delta_\text{max}>\pi/2$ to create a monopole with a single link update. This means that for
$\pi/2>\delta_\text{max}>\pi/3$ the single link update is not ergodic and cannot be used on its own.
To have an ergodic algorithm we need to update at least three faces of a cube at the
same time, which can only be done by updating more than one link at a time, as illustrated in Fig.~\ref{fig:monop_update}.
The minimal update to achieve this is shown in the lower part of Fig.~\ref{fig:monop_update} where two links of
a given plaquette are updated together. This update changes three plaquettes in each of the four cubes sharing the plaquette
common to the two updated links, and we thus have a chance to create four monopoles down to $\delta_\text{max}=\pi/3$ as required.

\section{The Helicity modulus}
The helicity modulus was first introduced in the $2d$ $XY$-model~\cite{Nelson:1977zz} where it quantifies the response
of the system to a twist in the boundary conditions. Because the twist is a boundary effect the helicity modulus is
an order parameter for a system with one massive (finite correlation length) and one massless (infinite correlation
length) phase. This is precisely the case of $4d$ lattice $U(1)$ gauge theory where the confining phase features massive
photons whereas they are massless in the Coulomb phase. In the context of a gauge theory the twisted boundary
conditions can also be thought of as an external electromagnetic flux~\cite{Vettorazzo:2003fg}.
More precisely, we define the helicity modulus as
\begin{equation}
  \label{eq:helmod}
  h \equiv \left.\frac{\partial^2f(\phi)}{\partial\phi^2}\right\rvert_{\phi=0},
\end{equation}
where $f$ is the free energy density in the presence of the external flux $\phi$. The flux is introduced by the replacement
\begin{equation}
  \label{eq:fluxadd}
  \cos(\theta_P)\to\cos(\theta_P+\phi)
\end{equation}
for all plaquettes in a given stack of plaquettes, i.e. all plaquettes in the set $\{P_{\mu\nu}(x)\mid \mu = \mu_0, \nu=\nu_0;x_\mu=x_0,x_\nu=y_0\}$.
The orientation and position of the pierced stack is arbitrary and with a suitable change of variables the flux can also be spread
out evenly over the $(\mu_0,\nu_0)$-planes. For the Wilson action $h$ is a simple difference of expectation values
\begin{equation}
  \label{eq:helmod_wil}
  h = \beta\left(\expv{\cos\theta_P}-\beta\expv{\left(\sum_\text{stack}\sin\theta_P\right)^2}\right),
\end{equation}
where the sum in the second term is over all plaquettes in the stack defined above. For the topological action on the other hand,
it is not possible to explicitly perform the derivatives. However, since the action for each configuration
is the same, the free energy is given solely by the entropy, i.e. by the number of configurations with a given
flux $\phi$. This can be measured by promoting the flux to a dynamical variable, which is updated along with the
link angles \cite{Bietenholz:2012ud}. By measuring the probability distribution $p(\phi)$ (via a histogram method for example) of the visited fluxes
one thus obtains the full $2\pi$ periodic free energy \cite{Vettorazzo:2003fg} and the helicity modulus
\begin{equation}
  \label{eq:helmod_top}
  h = -\left.\frac{\partial^2\log p(\phi)}{\partial\phi^2}\right\rvert_{\phi=0}.
\end{equation}
Alternatively, and more accurately, one can use all the global information from $p(\phi)=e^{-f(\phi)}$ and fit it to the classical
ansatz~\cite{Vettorazzo:2003fg}
\begin{align}
  \label{eq:helmod_ana}
  f(\phi) &= -\log\sum_ke^{-\frac{\beta_R}{2}\left(\phi-2\pi k\right)^2}\nonumber\\
  &=-\log\vartheta_3\left(\frac{\phi}{2};e^{-\frac{1}{2\beta_R}}\right) - \frac{1}{2}\log 2\pi\beta_R,
\end{align}
where $\beta_R$ plays the role of the renormalized coupling in the Coulomb phase and $\vartheta_3(z,q)$ is a Jacobi theta function.
From this ansatz we can extract the curvature at $\phi=0$, i.e. $h$, analytically and we thus obtain both the helicity modulus and the
renormalized coupling at the same time.
We further note that they approach each other exponentially fast for large $\beta_R$. Together with eq.~\eqref{eq:helmod_wil} we
see that this means that $\beta_R\approx h \to \beta-1/4$ as $\beta\to\infty$, which is to say that the coupling constant
is not renormalized in the continuum limit which is of course common knowledge.

\section{Continuum limit}
It is important to dwell a little on the matter of a continuum limit for the topological action. Since all the plaquettes
are forced to unity when $\delta_\text{max}\to0$ one expects that in this limit the correlation length diverges and thus
that it defines a continuum limit. This point of view was examined more thoroughly by Budczies and Zirnbauer in~\cite{Budczies:2003za}.
These authors consider a general weight function $w_t(U_P)$, which is a function of a plaquette variable $U_P$ and some parameter
(coupling) $t$. Granted that there exists a $t_c$ such that $w_{t_c}(U_P)=\delta(U_p-\text{id})$ and that for $t\neq t_c$ the weight
function is some smeared version of the $\delta$-function, then the lattice gauge theory with partition function
\begin{equation}
  Z_t = \int\rd[U]\,\prod_Pw_t(U_P)
\end{equation}
has a continuum limit as $t\to t_c$. Furthermore, the authors claim that under ``\emph{favorable conditions}'', the continuum theory
will be Yang-Mills theory. It is not precisely defined what conditions are considered favorable, but close to the identity element,
the plaquette variable is well approximated by $U_P=e^{ia^2F_P}\approx 1+ia^2F_P-a^4F_P^2$. Thus, in order for the continuum action
to be $\propto \int\text{Tr}F^2$ the weight function $w_t$ certainly has to satisfy some conditions on the moments of the tangent vectors of
the Lie group. At the very least the first moment must vanish and the second moment needs to exist and have the correct sign.
The authors indeed give an example in~\cite{Budczies:2003za} of a weight function, in two dimensions and for gauge group U$(N)$,
which satisfies the $\delta$-function condition but which has the wrong continuum limit. The problem is identified with the
non-existence of the second moment for the considered weight function.

The topological action which we use clearly satisfies the $\delta$-function constraint since the weight function has support only on
a compact region of width $\pm\delta_\text{max}$ around the identity element and thus goes to $\delta(U_P-\text{id})$ as $\delta_\text{max}\to0$.
Because of the compact support and invariance under Hermitian conjugation we also conclude that the first moment vanishes and that the
second is positive as it should. It is therefore probable that this action will have the correct quantum continuum limit and indeed all
numerical evidence suggests that it does.

A simple check one can perform is to use for $w_t(U_P)$ a combination of angle restriction and Wilson plaquette term with negative $\beta$. By taking
$\delta_\text{max}\to0$ the action still satisfies the $\delta$-function constraint but the negative value of $\beta$ will try to bend
the distribution in the wrong direction to make the second moment of $w_t$ negative. Clearly, for a fixed value of $\beta$ the action will still
be almost flat as long as $\delta_\text{max}$ is small enough, so in order to change the continuum limit, $\beta$ needs to be taken to
$-\infty$ at the same time as $\delta_\text{max}\to0$. Then, if the magnitude of $\beta$ is large enough we expect that the continuum limit
is spoiled. This can also be observed in numerical simulations, and although it is somewhat of a pathological example it still gives some
insight as to when one can expect to obtain the correct continuum limit.

\section{Free energy}

Here, we show how to evaluate the free energy, analytically in a $1d$ toy
model, and numerically for more realistic cases.

\subsection{$1d$ $XY$ model}
Consider a periodic chain of $N$ spins $s_i\in{}O(2)$ with a topological action which restricts the angle of each
link $\ell_i=s_is^\dagger_{i+1}$ to be smaller than $\delta_\text{max}$.
Let $\ell_i = \exp(i\theta_i),\; \theta_i\in[-\delta_\text{max},\delta_\text{max}]$. The partition function of this model then
takes a very simple form,
\begin{equation}
  Z = \int\limits_{-\delta_\text{max}}^{\delta_\text{max}}\prod_{i=1}^N\frac{\rd \theta_i}{2\delta_\text{max}}\delta\left(\exp\left(i\sum_{i=1}^N\theta_i\right)-1\right)
\end{equation}
and describes a collection of $N$ non-interacting, constrained links with the only condition that the product of all links is one.
The normalization of the angle integrals serves to keep $Z$ finite as the number of links is taken to infinity and is just a subtraction of the
ground state energy.

The total angle can take values $2\pi m,\;m\in \{-\left\lfloor\frac{N\delta_\text{max}}{2\pi}\right\rfloor,\ldots,
\left\lfloor\frac{N\delta_\text{max}}{2\pi}\right\rfloor\}$ and thus $m$ is the winding number or topological charge of the
system. The partition function can be expressed solely in terms of the total angle by convoluting the uniform distributions
of the individual links $N$ times. The distribution of the sum of $N$ i.i.d. uniform variables converges very rapidly
to the normal distribution, in this case with zero mean and variance $N\delta_\text{max}^2/3$. Anticipating the $N\to\infty$
limit we thus neglect the small deviations from the normal distribution and write
\begin{align}
  Z &= \sqrt{\frac{3}{2\pi N\delta_\text{max}^2}}\int\limits_{-N\delta_\text{max}}^{N\delta_\text{max}}\!\!\!\!\!\!\!\!\rd \theta\,\exp\left(-\frac{3\theta^2}{2N\delta_\text{max}^2}\right)\delta\left(\exp\left(i\theta\right)-1\right)\nonumber\\
  &= \frac{\sqrt{3}}{(2\pi)^{3/2}\alpha}\sum_{m=-\left\lfloor\sqrt{N}\alpha\right\rfloor}^{\left\lfloor\sqrt{N}\alpha\right\rfloor}\exp\left(-\frac{3}{2}\left(\frac{m}{\alpha}\right)^2\right),
\end{align}
where we have defined $\alpha\equiv\frac{\sqrt{N}\delta_\text{max}}{2\pi}$. We can now take $N\to\infty$ whilst keeping
$\alpha$ fixed to obtain
\begin{align}
  Z &= \frac{\sqrt{3}}{(2\pi)^{3/2}\alpha}\sum_{m=-\infty}^\infty\exp\left(-\frac{3}{2}\left(\frac{m}{\alpha}\right)^2\right)\nonumber\\
  &= \frac{\sqrt{3}}{(2\pi)^{3/2}\alpha}\vartheta_3\left(\exp\left(-\frac{3}{2\alpha^2}\right)\right),
\end{align}
where $\vartheta_3(q)\equiv\vartheta_3(0,q)$ is the third Jacobi elliptic theta function.
Since the sum in the partition function is over the winding number $m$ it is straightforward to calculate $\expv{m^2}$
and the topological susceptibility $\chi_t = \frac{1}{\beta}\expv{m^2}$. In the limit $\beta=Na\propto\alpha^2\to\infty$ (where $a$ is the lattice spacing) one should find
$\chi_t = \frac{1}{4\pi^2I}$ where $I$ is the moment of inertia of the quantum rotor which the model describes.
This allows us to determine $\alpha$ in terms of $\beta$ and $I$ and the result is $\alpha=\frac{\sqrt{3\beta/I}}{2\pi}$
which leads to
\begin{equation}
  Z = \sqrt{\frac{I}{2\pi\beta}}\vartheta_3\left(\exp\left(-\frac{2\pi^2I}{\beta}\right)\right).
\end{equation}
With Poisson's summation formula we can go from the winding number representation to the energy representation in which
\begin{equation}
  Z = \vartheta_3\left(\exp\left(-\frac{\beta}{2I}\right)\right) = \sum_{k=-\infty}^\infty\exp\left(-\frac{k^2\beta}{2I}\right).
\end{equation}
It is now evident that the excited states are doubly degenerate and the energy differences are $E_k-E_0 = \frac{k^2}{2I}$
as is well known. The topological susceptibility is given in the two representations by
\begin{align}
  \chi_t &= \frac{\exp\left(-\frac{2\pi^2I}{\beta}\right)\vartheta'_3\left(\exp\left(-\frac{2\pi^2I}{\beta}\right)\right)}
{\beta\vartheta_3\left(\exp\left(-\frac{2\pi^2I}{\beta}\right)\right)}\nonumber\\
&= \frac{1}{4\pi^2I}\left(1-\beta\frac{\exp\left(-\frac{\beta}{2I}\right)\vartheta'_3\left(\exp\left(-\frac{\beta}{2I}\right)\right)}
{\vartheta_3\left(\exp\left(-\frac{\beta}{2I}\right)\right)}\right).
\end{align}
Since the elliptic function and its derivative are analytic functions  $\forall\beta \in \R^+$ there is no phase transition
but there are two distinct regimes with a rather abrupt crossover. In the low temperature regime, $\beta/I\gtrsim 10$,
the partition function is almost independent of $\beta$ and the topological susceptibility is very close to its zero
temperature value $(4\pi^2I)^{-1}$ whereas in the high temperature region, $\beta/I \lesssim 10$, the partition function is
approximately $\sqrt{2\pi\beta}$ and $\chi_t$ rapidly drops to zero.

Note that, when $N\delta_\text{max}<2\pi$, topological excitations are forbidden and $\chi_t=0$.
However, the continuum limit is obtained while keeping $(N\delta_\text{max}^2$ fixed,
so that the lattice spacing varies $\propto\delta_\text{max}^2$.
Therefore, in this $1d$ model the parameter region where $\chi_t=0$ disappears in the continuum limit.

\subsection{Higher dimensions and gauge theories}
\begin{figure}[htpb]
\centering
\includegraphics[width=\linewidth]{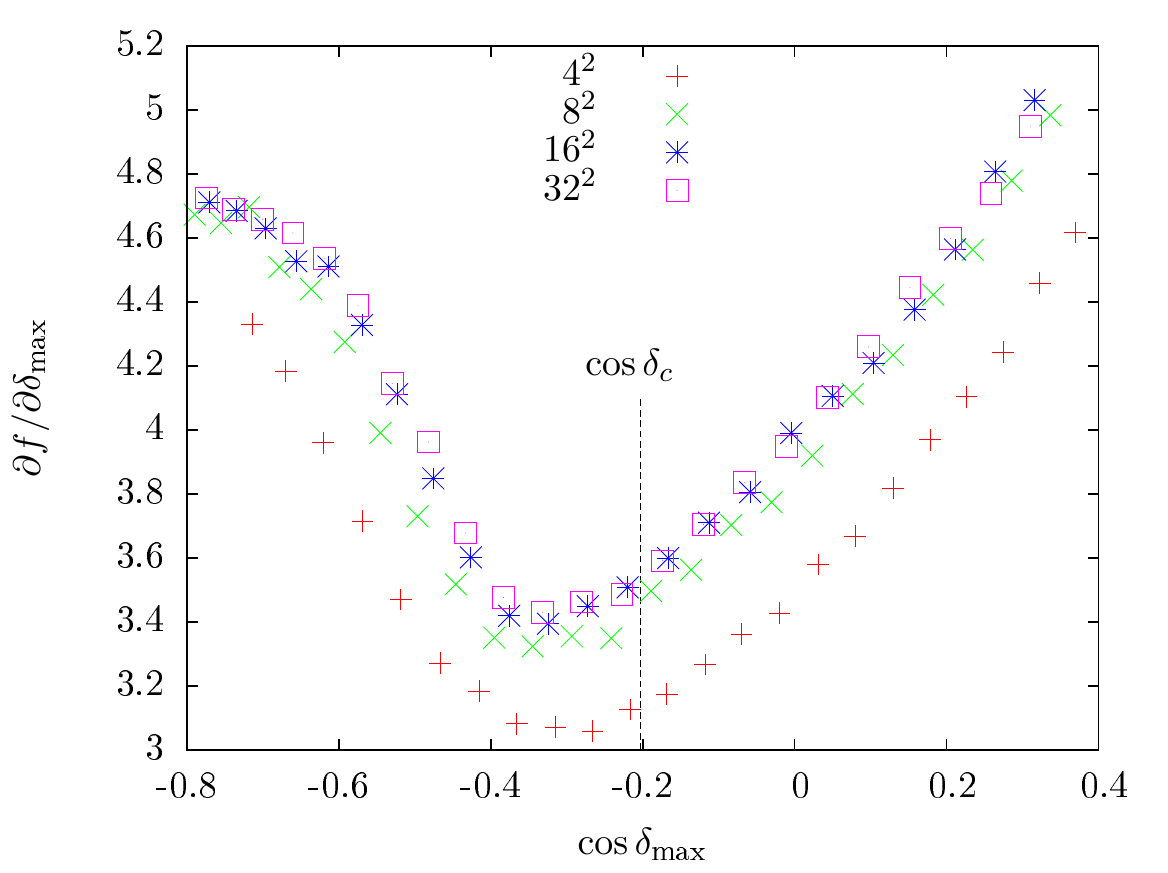}\\
\includegraphics[width=\linewidth]{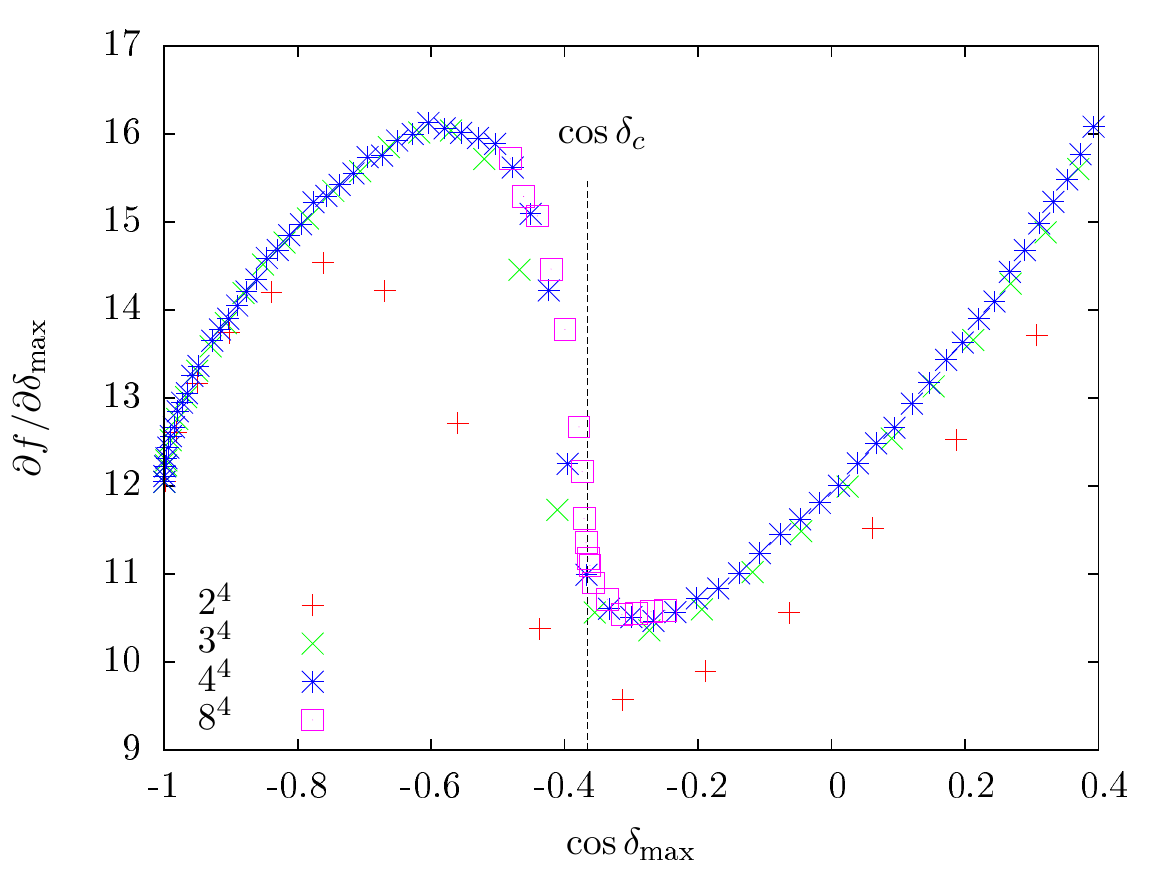}
\caption{The derivative $\partial f/\partial\delta_\text{max}$ of the free energy density $f = -V^{-1}\log Z$ for the $2d$ $XY$-model (\emph{upper panel}) and the $4d$ $U(1)$ gauge theory
(\emph{lower panel}) as obtained from Monte Carlo simulations with a topological action. There is a clear distinction
between the smooth derivative in the $XY$-model which has an infinite
order phase transition and the discontinuous behavior, signaling a first order transition, in the $U(1)$ gauge theory.
The vertical line marks the critical restriction $\cos\delta_c$. In the
case
of the XY-model it has been taken from~\cite{Bietenholz:2012ud} where it
was extracted from a fit of the diverging correlation length.
}
\label{fig:free_en}
\end{figure}

In higher dimensions, due to the lattice Bianchi identities, the integration over the constrained variables no longer
factorizes and we can not calculate the partition function analytically anymore. However, in the small $\delta_\text{max}$
regime where there are no topological defects the partition function must be
\begin{equation}
  Z = \left(2\delta_\text{max}\right)^{n_\text{d.o.f}}
\end{equation}
(or one, depending on the normalization), where $n_\text{d.o.f}$ is the number of independent degrees of freedom.
As the topological defects are turned on, the functional dependence on $\delta_\text{max}$ will change and there will
be a high order and practically undetectable phase transition. As $\delta_\text{max}$ is further increased the topological
defects will start to play a more important role and eventually the \emph{real} phase transition of the model will occur.
If one would have access to the partition function, or free energy, one could directly extract the properties of the transition.
Fortunately, since the topological action is constant, the partition function is pure entropy and can thus be measured
by Monte Carlo simulations by simply counting the number of configurations at a given value of $\delta_\text{max}$.
If Figure \ref{fig:free_en} we show the derivative of the free energy density $f = -V^{-1}\log Z$ with respect to $\delta_\text{max}$ for the 2$d$ $XY$-model
(\emph{upper panel}) and the $4d$ $U(1)$ gauge theory (\emph{lower panel}) for various lattice volumes, obtained by Monte Carlo
simulations. It is clear that the derivative
is smooth in the $XY$-model where the transition is of infinite order (BKT) and that it is discontinuous in the $U(1)$ case where
the transition is first order.

\section{Results}
\begin{figure}[htpb]
\centering
\includegraphics[width=\linewidth]{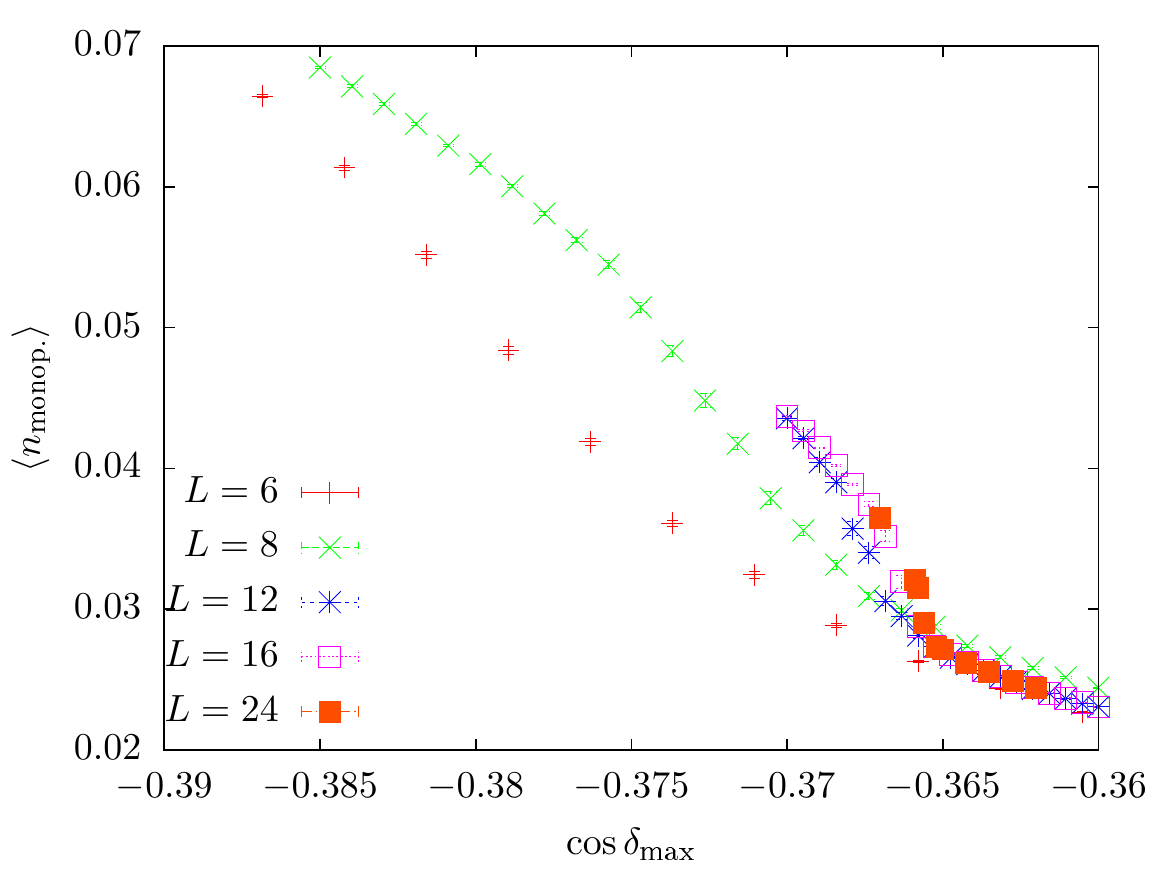}\\
\includegraphics[width=\linewidth]{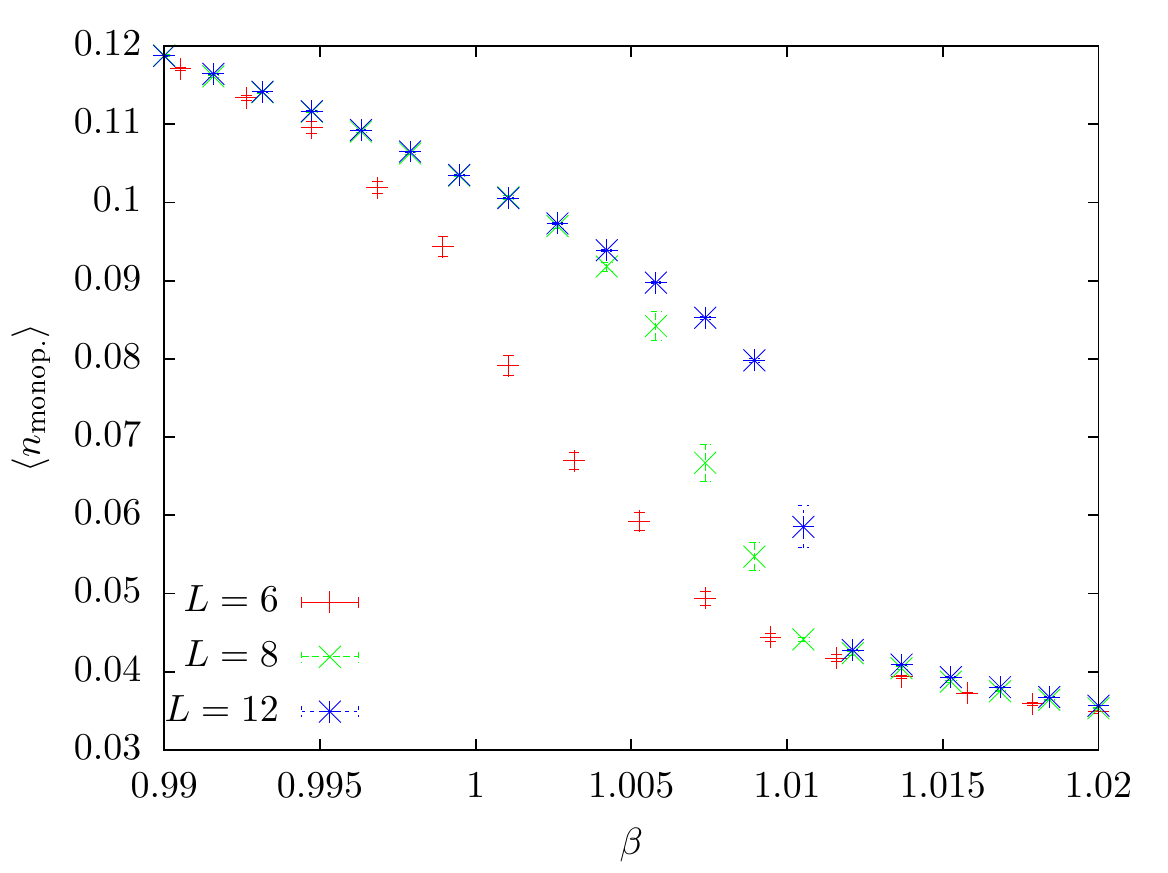}
\caption{The monopole density $n$ for the topological action (\emph{upper panel}) and the Wilson action
(\emph{lower panel}). For the Wilson action the first order nature of the transition is rather evident
even for a $12^4$ lattice whereas for the topological action we have to go to much larger lattices to see
a fairly distinct jump.}
\label{fig:dens}
\end{figure}

\begin{figure}[htpb]
\centering
\includegraphics[width=\linewidth]{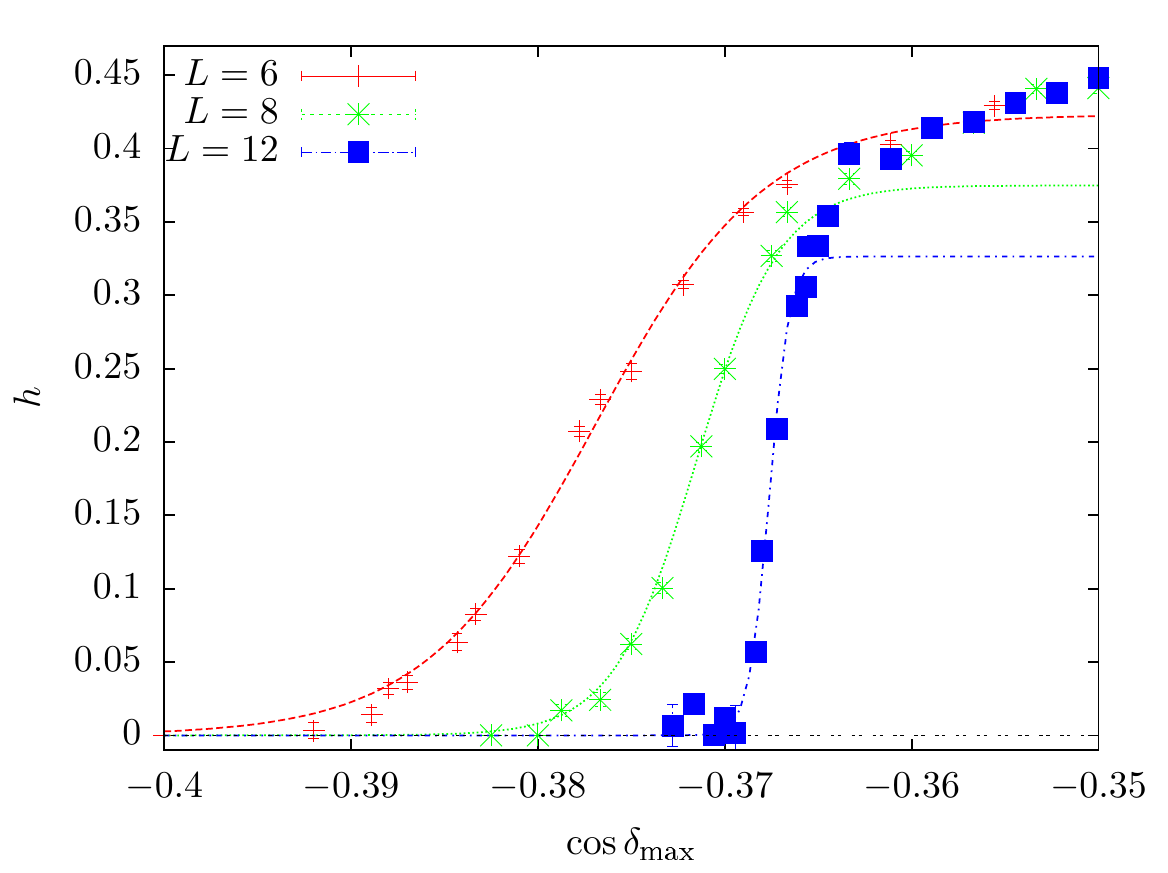}\\
\includegraphics[width=\linewidth]{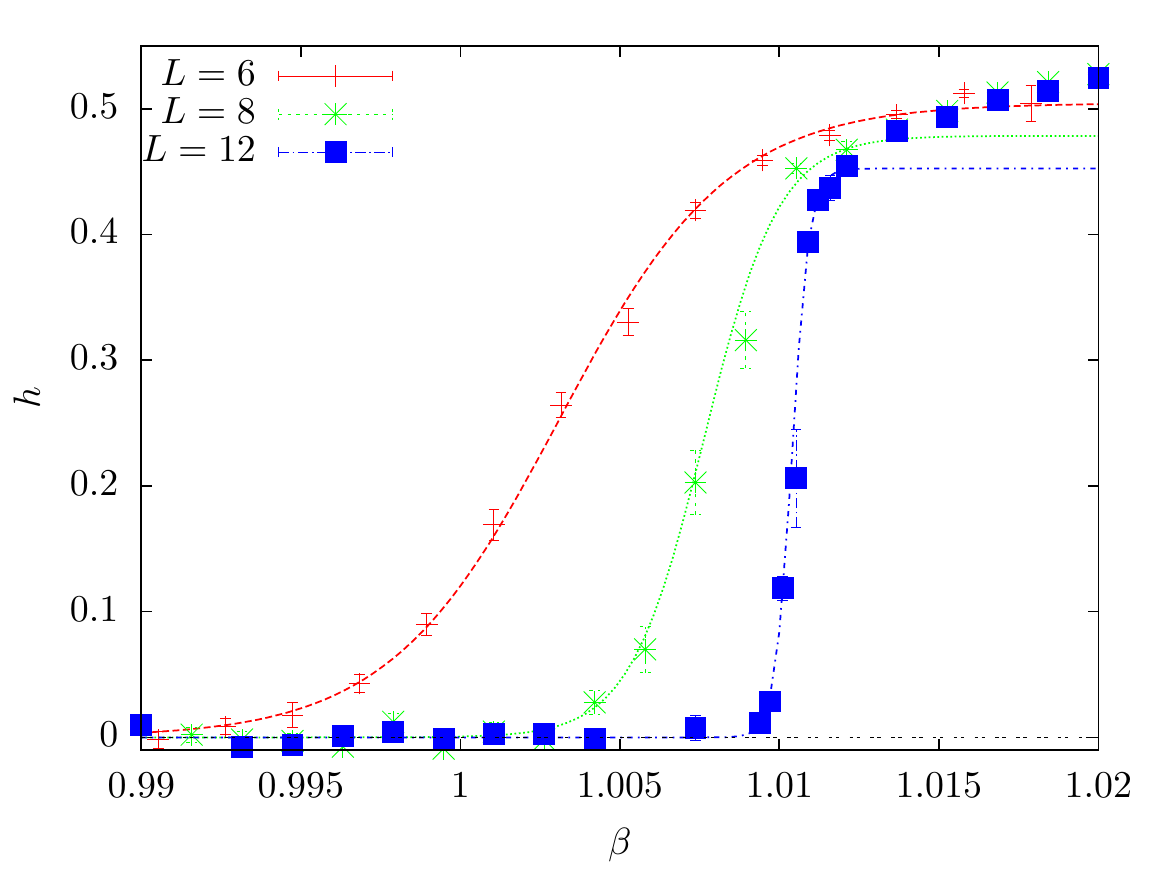}
\caption{The helicity modulus $h$ for the topological action (\emph{upper panel}) and the Wilson action
(\emph{lower panel}). The lines are the best fit to eq.~\eqref{eq:hel_exp}, which describes the data in the
confining phase (the model assumes a constant $h$ in the Coulomb phase) very well for both actions.}
\label{fig:helmod}
\end{figure}

Let us now turn to the numerical results. Primarily what we are interested in is the phase structure of the model and the
order of the possible deconfinement transition. To this end we have measured the monopole
density and the helicity modulus as a function of the restriction $\cos\delta_\text{max}$. We compare these results with
the corresponding observables obtained with the Wilson action in Figs.~\ref{fig:dens} and \ref{fig:helmod}:
it is obvious that the transition is even weaker than the weak first order transition seen with the Wilson action.
We can try to quantify the strength of the transition by fitting the helicity modulus in the confining phase using a simple
model of a first order transition~\cite{Borgs:1990xs,Vettorazzo:2003fg}.
\begin{equation}
  \label{eq:hel_exp}
  h(x) = \frac{h_+}{1+X^{-1}\exp\left(-V\Delta{}f(x-x_c)\right)},
\end{equation}
where $h_+$ is the helicity modulus in the Coulomb phase (which is assumed to be constant), $\Delta{}f$ is the latent heat,
$X$ is an anisotropy factor between the two phases and $x$ is the coupling, either $\beta$ or $\cos\delta_{\rm max}$. After taking
finite size effects into account the best fit is shown as the lines in Fig.~\ref{fig:helmod}. The data is
well described by the ansatz and one finds that the fitted value of the latent heat for the Topological action is about half
of what it is using the Wilson action, which is consistent with the weaker transition seen in the monopole density.

\begin{figure}[htpb]
\centering
\includegraphics[width=\linewidth]{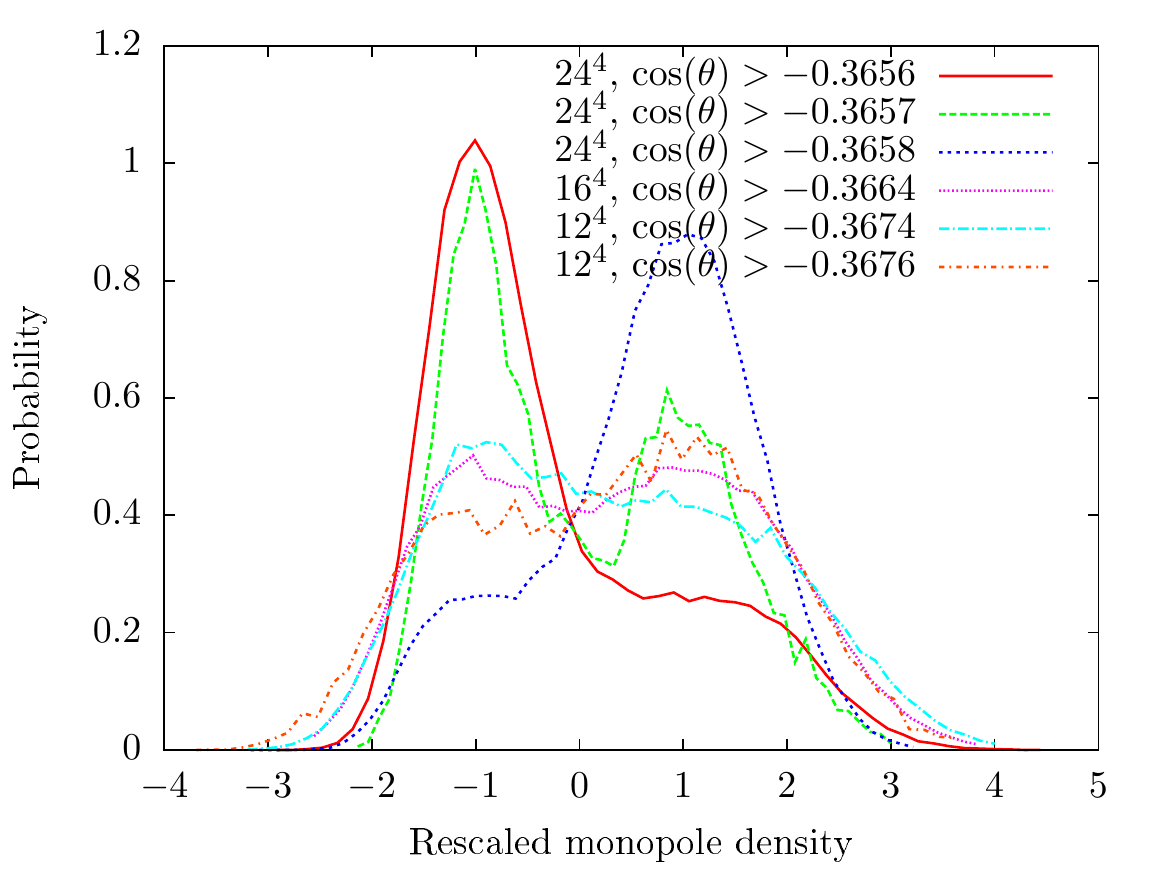}\\
\includegraphics[width=\linewidth]{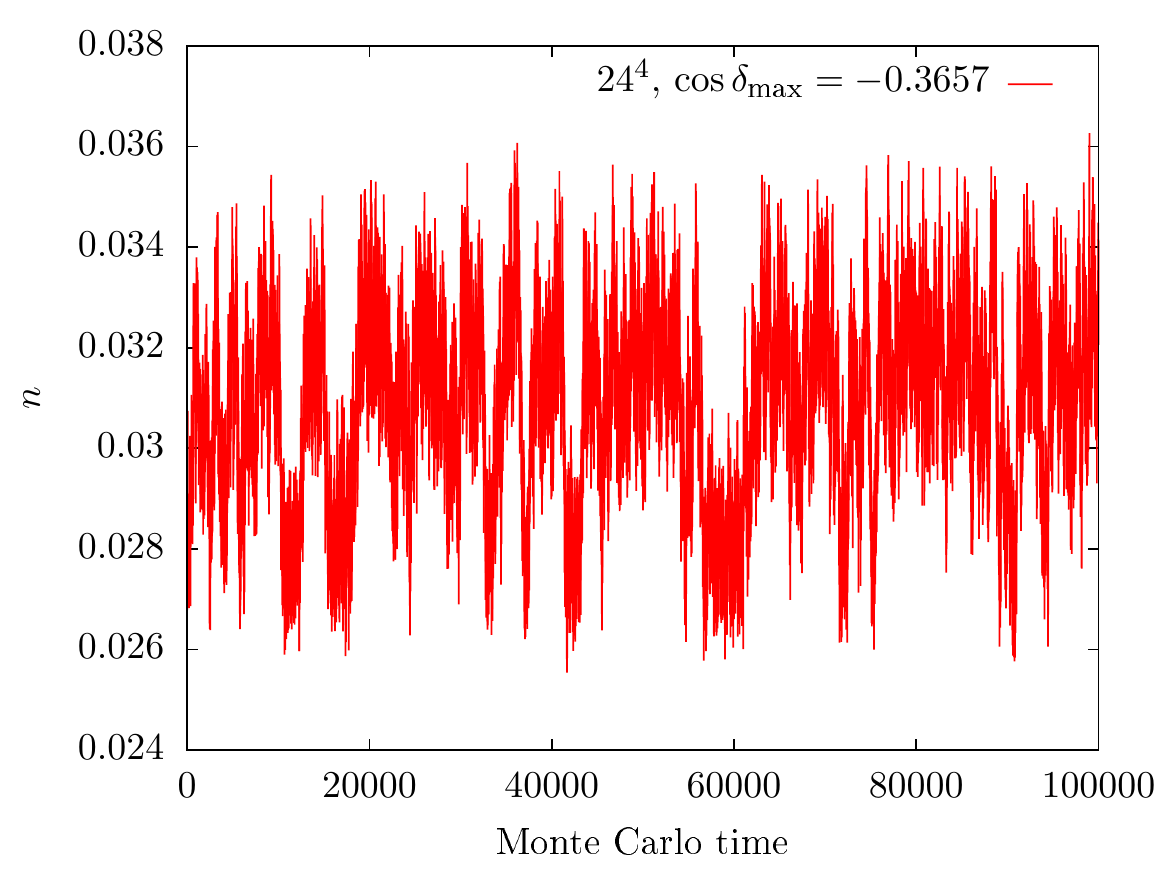}
\caption{The probability distribution of the monopole density close to the transition for various volumes 
(\emph{upper panel}) and the corresponding Monte Carlo history for the $24^4$ volume (\emph{lower panel}).
The distributions are rescaled in such a way that one peak is at -1 and the other at +1. That way the separation
of the peaks in relation to the widths can be directly compared between different volumes. It is evident that
the two peaks become more distinct for larger volumes which indicates a first order transition. Also the obvious
tunneling between two different states in the Monte Carlo history backs up this statement.}
\label{fig:dens_hist}
\end{figure}

To further establish that the transition really is first order we show histograms of the monopole density close to the transition
for three different volumes in Fig.~\ref{fig:dens_hist}. A double peak structure is formed and enhanced as the volume
increases, which is a clear indication that the transition is first order. Also the Monte Carlo history shows clear
tunneling events between two metastable states. Together with the discontinuity in the first derivative of the free
energy with respect to the cutoff we conclude that the topological action has a first order transition at
$\delta_\text{max} \approx 1.95$.

\begin{figure}[htpb]
\centering
\includegraphics[width=\linewidth]{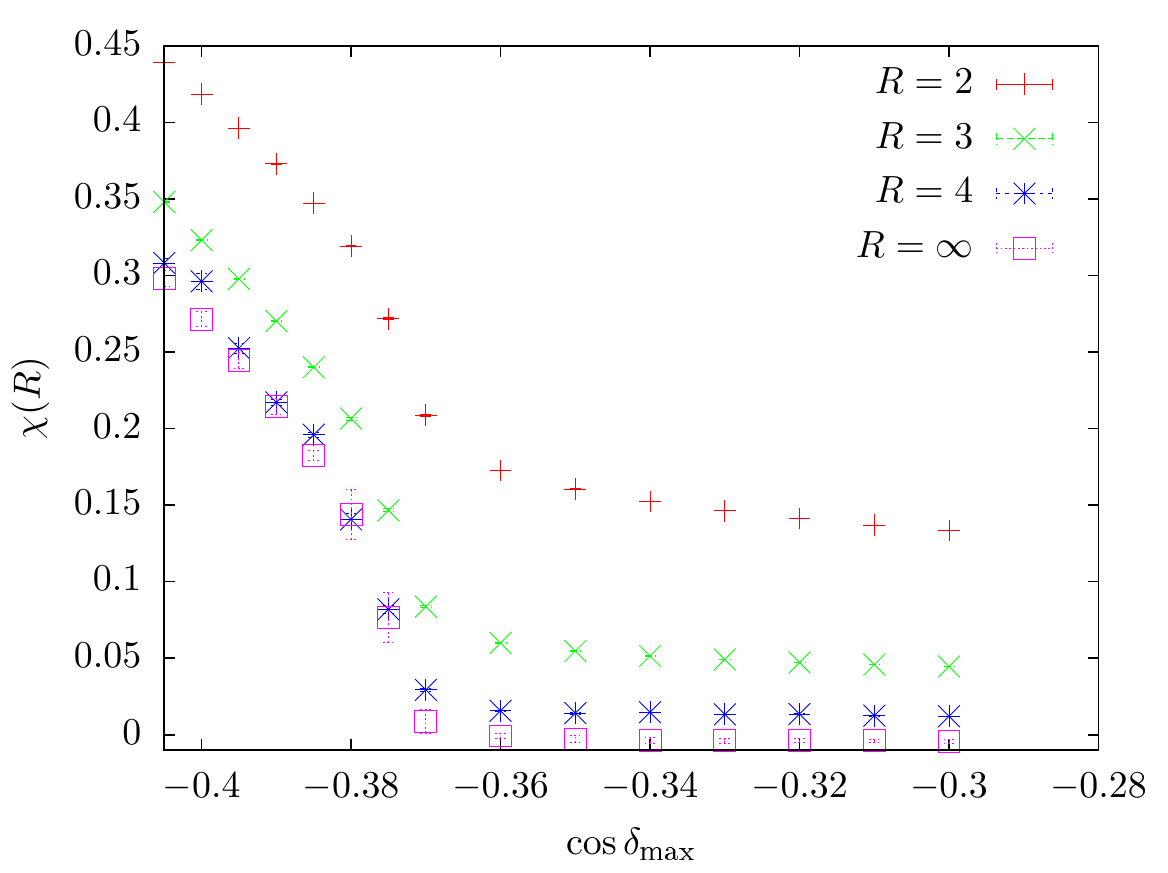}
\caption{The Creutz ratios $\chi(R)$ given by eq.~\eqref{eq:cr} as well as an $R\to\infty$ extrapolation assuming corrections of the
form $e^{-R}$ as a function of the restriction $\cos\delta_\text{max}$ obtained on an $8^4$ lattice. There is a clear transition between
a confining phase with nonzero string tension and a deconfined phase with an perimeter law for the Wilson loops.}
\label{fig:cr}
\end{figure}

To determine the characteristics of the two phases we look at how Wilson loops of different sizes behave. Naively,
we expect an area law when $\delta_\text{max}$ is close to $\pi$ since the interaction between plaquettes will be
very weak, as for the Wilson action where $\beta\ll 1$. This can be seen in the left panel of Fig.~\ref{fig:allowed_angles}. If the forbidden regions become
very narrow then the individual links are hardly influenced by their neighbors and each plaquette angle is more or less
uniformly distributed in the interval $[-\delta_\text{max},\delta_\text{max}]$ which gives an average plaquette trace of
$\sin(\delta_\text{max})/\delta_\text{max}$. For a loop with area $A$, this is raised to the $A$'th power. For restrictions $\delta_\text{max}$
close to zero on the other hand, the links are heavily influenced by their neighbors (right panel of Fig.~\ref{fig:allowed_angles})
and the total angle of the loop should depend on the perimeter rather than the area. This is demonstrated in Fig.~\ref{fig:cr} where we
show the Creutz ratios
\begin{equation}
  \label{eq:cr}
  \chi(R) = -\log\frac{\expv{W(R,R)}\expv{W(R-1,R-1)}}{\expv{W(R,R-1)}\expv{W(R-1,R)}},
\end{equation}
where $W(I,J)$ is a planar, rectangular Wilson loop with sides $I$ and $J$. We have performed the $R\to\infty$ extrapolation under the
assumption that the corrections are of the form $e^{-R}$. Note that this is not a precise measurement of the string tension but rather
a characterization of the two phases. We have also checked that the magnitude of the Polyakov loop acquires a vacuum expectation value
in the low monopole density phase.

\begin{figure}[htpb]
\centering
\includegraphics[width=\linewidth]{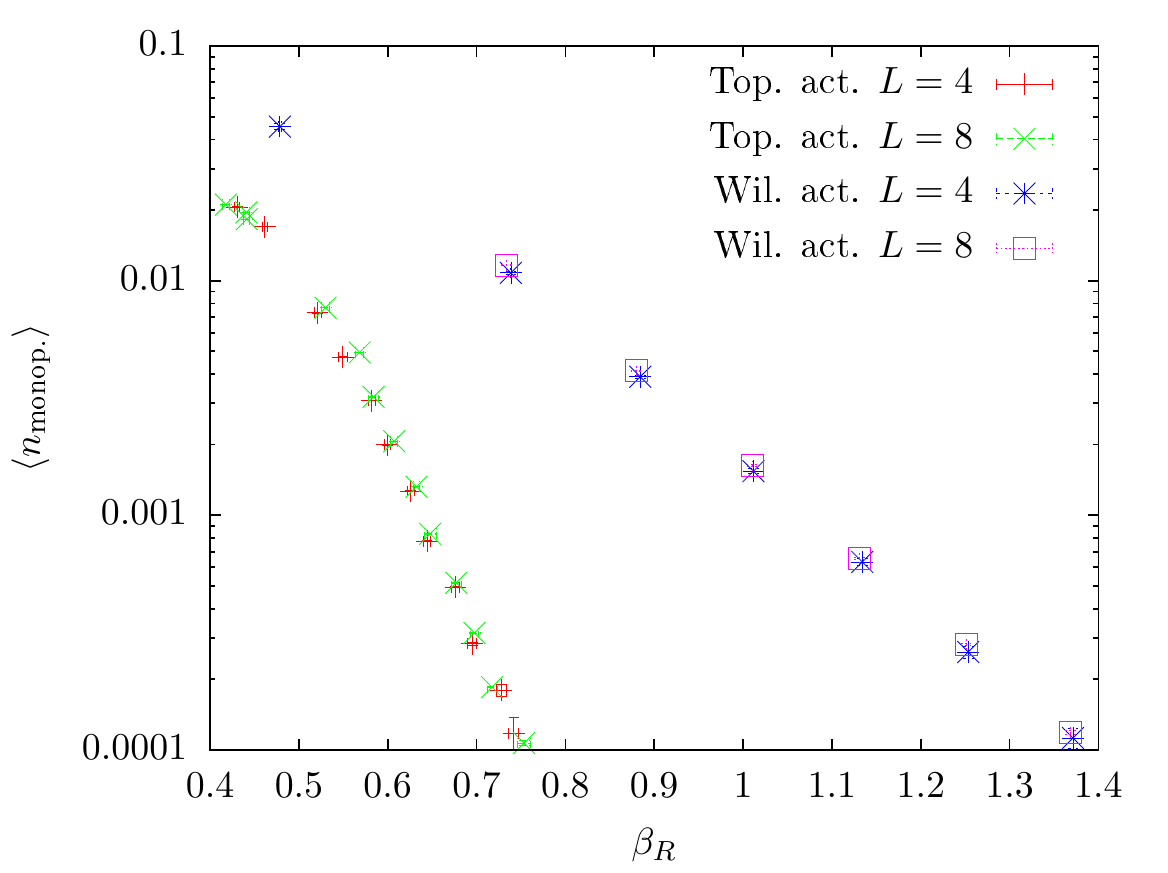}
\caption{The monopole density in the Coulomb phase as a function of the renormalized coupling $\beta_R$, for the topological and the Wilson actions. 
The different rates
of decay could be attributed to different lattice spacings for the two actions.}
\label{fig:dens_helicity}
\end{figure}

Another interesting thing to investigate is how the monopole density depends on the renormalized coupling. The monopole mass
is proportional to $\beta_R = e_R^{-2}$ and the density decreases exponentially with the mass. This is a statement about physics
so it gives us a direct way to compare the two actions. In Fig.~\ref{fig:dens_helicity} we show the monopole density as a function
of the renormalized coupling and we see a clear exponential decay as expected. For the topological action the decay is significantly
faster, which could be interpreted as a reduction in the discretization errors: for a given effective coupling, there are fewer
lattice artifacts (monopoles) that disturb the order of the system. For $\delta_\text{max}<\pi/3$ the density is even strictly zero
and the model is completely insensitive (up to trivial rescalings) to further reduction of $\delta_\text{max}$.

\begin{figure}
\centering
\includegraphics[width=\linewidth]{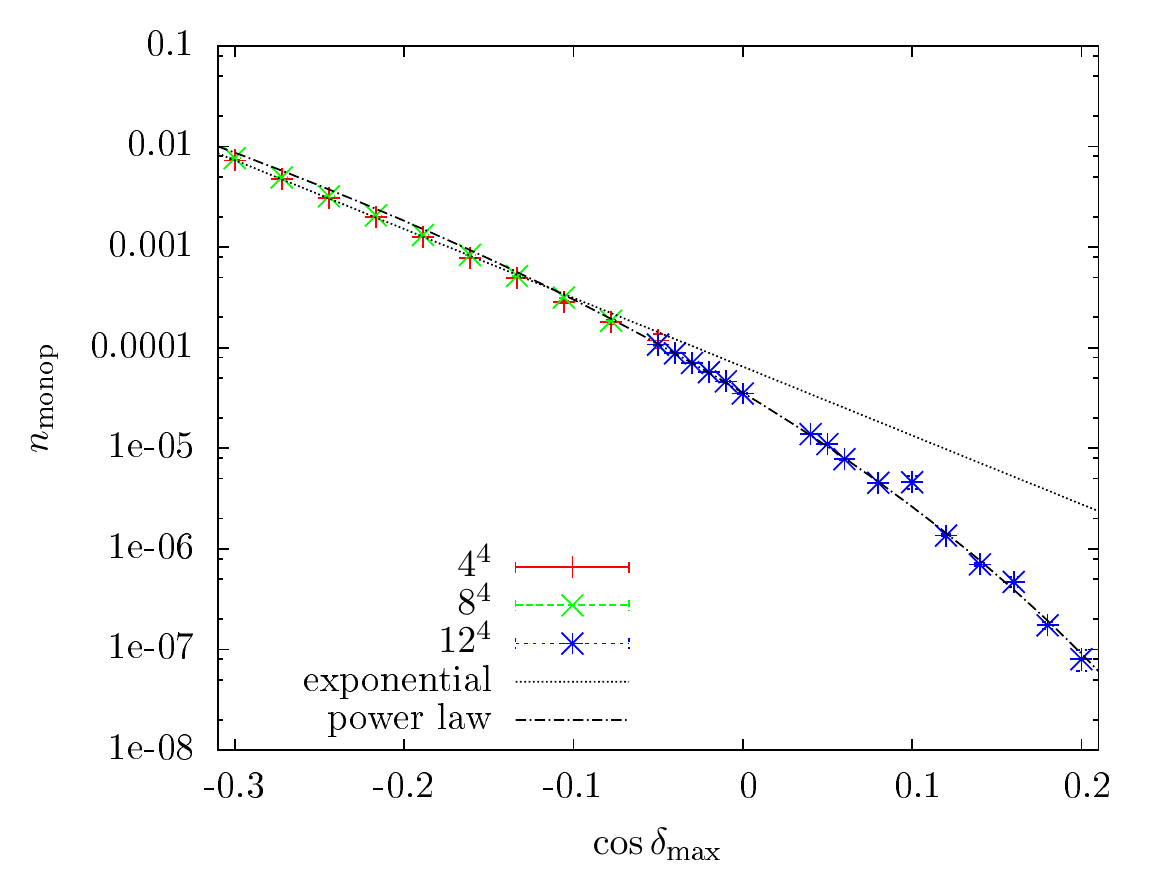}
\caption{The monopole density for the topological action as a function of the restriction. There seems to be a smooth transition from
an exponential decay to a power law at $\cos\delta_\text{max}\approx -0.12$.}
\label{fig:nmonop_top}
\end{figure}

With a mix of single- and two-link updates we have been able to measure the monopole density down to densities around $10^{-8}$. The
exponential dependence on $\cos\delta_\text{max}$ persists to $\delta_\text{max}\approx 1.69$ after which the density smoothly changes
into a power law in $(1/2 - \cos\delta_\text{max})$ with an exponent which is fitted to be $11.70(6)$ as can be seen in Fig.~\ref{fig:nmonop_top}.
We tentatively ascribe this change of functional behavior to the approach
of a phase transition. A naive argument, which works well in the 2$d$ $XY$-model, leads to a monopole
density which is polynomial in the small deviation $(\delta_\text{max}-\pi/3)$. The argument is based on convolutions of (near) uniform plaquette
or link distributions. To create a single vortex in the spin model close to the threshold $\pi/2$ we need to convolve the link angle distribution four times, which makes
the joint distribution $\propto (4\delta_\text{max}-\theta)^3$
for the cumulative angle $\theta$ around a plaquette. 
This needs to be evaluated at $\theta=2\pi$ (one vortex) which gives
a vortex probability $\propto(\delta_\text{max}-\pi/2)^3$. Vortices always come in pairs so we expect that the density is proportional
to $(\delta_\text{max}-\pi/2)^6$ which is in good agreement with what we have obtained from Monte Carlo simulations. 
By a similar argument one would expect a monopole
density $\propto(\delta_\text{max}-\pi/3)^{20}$ due to six plaquettes in 4 cubes containing a monopole. The deviation in the power law from the predicted 
$20$ to the observed $\approx12$ is rather large, but the argument does not take into account that the 4 monopoles are not independent of each other, so it
is not so surprising that one finds a smaller exponent.

\section{Conclusions}
We have simulated $U(1)$ lattice gauge theory using an unconventional ``topological'' action. We find that this action describe
the same physics as the Wilson action, i.e. there is a confining strong coupling phase where magnetic monopoles condense and
Wilson loops follow an area law, separated by a (weak) first order transition from a Coulomb phase with an exponentially suppressed
monopole density and a perimeter law for the Wilson loops. We have, in this specific case, not found any concrete advantages which would
motivate the choice of this action over the Wilson action although at a given value of the effective coupling in the Coulomb phase
there are significantly fewer monopoles (lattice artifacts). This is in line with other known cases where a topological action reduces discretization
errors~\cite{Bietenholz:2012ud}. Perhaps the most interesting approach is to search for optimized combinations of a standard action and constrained fields.
An interesting feature of the topological action is the direct access to the free energy itself.

One interesting open question is the nature of the extra transition at $\delta_\text{max}=\pi/3$ where there is a non-analyticity in
the monopole density as it goes from nonzero to strictly zero. 
A similar phenomenon occurs at $\delta_\text{max}=\pi/2$ for an XY model,
and when plaquettes become restricted by the ``admissibility condition'' in
gauge theories.
One may argue, however, in the $U(1)$ case at least, that this transition will have no impact on the physics
because the monopole density close to the transition is extremely small anyway.

\section*{Acknowledgments}
We thank Michele Pepe for useful discussions.

\appendix

\bibliography{top_action.bib}

\end{document}